# Real-time Data-driven Quality Assessment for Continuous Manufacturing of Carbon Nanotube Buckypaper

Xinran Shi[1], Xiaowei Yue[2*], Zhiyong Liang[3], and Jianjun Shi[1]

[1]H. Milton Stewart School of Industrial and Systems Engineering Georgia Institute of Technology;
[2]Grado Department of Industrial and Systems Engineering, Virginia Tech,
[3]High-Performance Materials Institute, Florida State University

*Abstract*—Carbon nanotube (CNT) thin sheet, or buckypaper, has shown great potential as a multifunctional platform material due to its desirable properties, including its lightweight nature, high mechanical properties, and good conductivity. However, their mass adoption and applications by industry have run into significant bottlenecks because of large variability and uncertainty in quality during fabrication. There is an urgent demand to produce high-quality, high-performance buckypaper at an industrial scale. Raman spectroscopy provides detailed nanostructure information within seconds, and the obtained spectra can be decomposed into multiple effects associated with diverse quality characteristics of buckypaper. However, the decomposed effects are high-dimensional, and a systematic quantification method for buckypaper quality assessment has been lacking. In this paper, we propose a real-time data-driven quality assessment method, which fills in the blank of quantifying the quality for continuous manufacturing processes of CNT buckypaper. The composite indices derived from the proposed method are developed by analyzing in-line Raman spectroscopy sensing data. Weighted cross-correlation and maximum margin clustering are used to fuse the fixed effects into an inconsistency index to monitor the long-term mean shift of the process and to fuse the normal effects into a uniformity index to monitor the within-sample normality. Those individual quality indices are then combined into a composite index to reflect the overall quality of buckypaper. A case study indicates that our proposed approach can determine the quality rank for ten samples, and can provide quantitative quality indices for single-walled carbon nanotube buckypaper after acid processing or functionalization. The quality assessment results are consistent with evaluations from the experienced engineers.

## I. INTRODUCTION

**B**uckypaper is a thin sheet made from an aggregate of carbon nanotubes, which could potentially provide high tensile strength, electrical and thermal conductivity, and optical properties [1]. Researchers showed that carbon nanotube (CNT) buckypaper has great application potential as a superb multifunctional platform material with functions ranging from heavy-duty materials to electronic circuits protector to artificial muscles [2], [3]. However, the mass adoption and applications of CNT buckypaper have experienced significant bottlenecks because of the high cost in production and large uncertainty in quality. A systematic real-time quality assessment of the high-performance buckypaper is urgently needed. The interests of the buckypaper characteristics include a specific type of multiwall carbon nanotube, geometric properties, width and diameter of the innermost wall, carbon unit cell ring size and connectivity, morphology, particle properties, and structural defects. Instead of studying these properties one-by-one, the macro perspective quality concerns of the CNT buckypaper include *consistency, uniformity,* and *defects*.

- *Consistency.* The degree of consistency indicates whether there is a gradual mean shift in the sequentially roll-to-roll fabrication process of CNT buckypaper.
- *Uniformity.* A sample is uniform if and only if the observations in the inspection area get similar features. The degree of uniformity reflects information such as the degree of alignment, the degree of functionalization, nanotube distribution, and dispersion.
- *Defects.* The within-sample defect information indicates whether there are defects in the CNT buckypaper. A specific band of Raman spectrum denotes corresponding defective information of the product.

To investigate the properties of the CNT buckypaper, various measurement tools are applied for characterization, including scanning electron microscopy (SEM), transmission electron microscopy (TEM), fast Fourier transform (FFT) of high-resolution TEM (HRTEM), Raman spectroscopy, and Fourier transform infrared spectroscopy (FTIR) [4]. People usually use SEM for morphology and dimension measurements, and purity quantification [5], and TEM and HRTEM for inner morphology measurements (including size, shape, purity, and disorder) [6]. FTIR spectroscopy can reflect the functionality of the product [7]. However, these techniques are not efficient, nor applicable for real-time quality monitoring during the continuous nanomanufacturing process. Raman spectroscopy attracts wide interest for its potential on providing rich nanostructure information about the purity, defects, buckypaper functionality, and nanotube alignment. The offline characterization methods based on Raman spectroscopy have been widely used in batch-to-batch nanomanufacturing of CNT buckypaper [8]-[11]. Although it has significant potential for quality monitoring [12]-[14], the real-time quality assessment based on in-line Raman spectroscopy is not well studied yet.

As in-line Raman spectroscopy is a nondestructive testing and provides detailed nanostructure information within seconds, we use it to collect real-time datasets for quality assessment of CNT buckypaper. Fig. 1 shows the in-line Raman inspection for 6-inch width roll-to-roll buckypaper production. For a sample zone in Fig. 1 (a), one collects SEM pictures (Fig. 1, b-d) from multiple sampling points for characterization, and inspects the corresponding Raman spectra (Fig. 1, e-g) for real-time quality assessment. The Raman peak intensity ratio of D-band and G-band ($I_D/I_G$) determines the alignment degree of the samples [12], [15] and structural defects to graphitization or crystallinity ratio [16], [17]. However, the intensity ratio cannot tell the detailed information about the product.

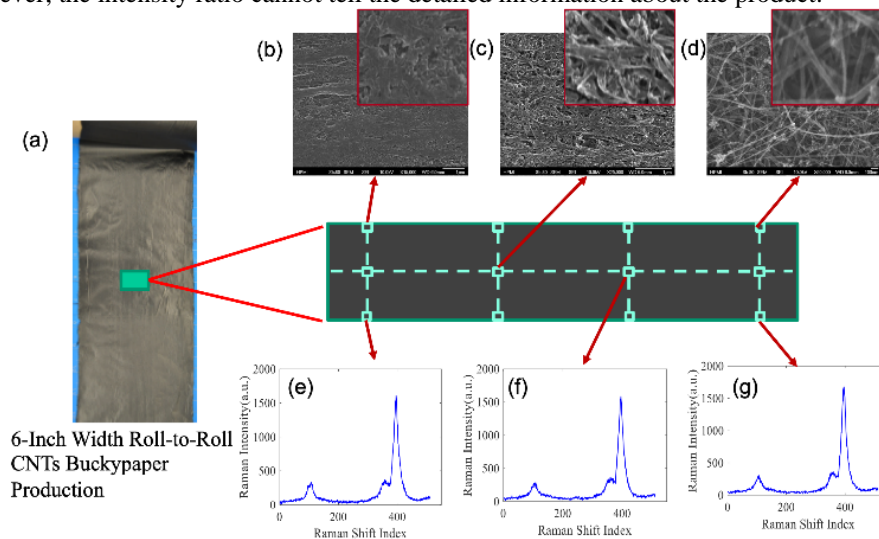

Fig. 1.  In-line Raman Spectroscopy Inspection for Buckypaper Production

From the SEM pictures (Fig. 1, b-d), one could see that the fabrication of CNT buckypaper is not consistent as the degree of alignment become worse along the fabrication process, and the degree of uniformity within each sample is different. Hence, the intensity ratio cannot be used as process assessment and quality control guidance directly. Furthermore, the moving speed of the roll-to-roll buckypaper fabrication process is a significant challenge for high signal-to-noise ratio (SNR) signal acquisition of in-line Raman spectroscopy, which may result in significant uncertainties in the intensity ratio. Therefore, we need to develop a systematic quantification method for real-time quality assessment of high-performance CNT buckypaper.

Yue et al. [18] used a generalized wavelet shrinkage method to increase the SNR of the in-line Raman spectra, which enables real-time quality control for CNT buckypaper manufacturing. Since the Raman spectra are collected from multiple channels and all the quality information, including consistency, uniformity, and defects are mixed in the datasets. They further developed a data decomposition approach, called wavelet-based penalized mixed-effects decomposition (PMD) [19], to obtain interpretable quality effects, i.e., (1) fixed effect that measures the fabrication consistency over time; (2) normal effects that reflect the uniformity of quality features within a sample; and (3) defective effects that indicate the existence and location of the defects in a sample area. A tensor mixed-effects model was also developed to separate fixed effects and random effects for high-dimensional arrays [20]. Although the quality features decomposed from the PMD are interpretable and correspond to multiple quality characteristics, it cannot be used to conduct the real-time evaluation for the product quality of CNT buckypaper directly. This is because the quality features decomposed from the PMD correspond to multiple high-dimensional parameters. Michael et al. proposed a quality assessment using X-ray photoelectron spectroscopy [21]. Horne and Liang mentioned using in-line

Raman spectroscopy for quality assessment [22]. However, they cannot obtain a unified and quantitative index for quality assessment of CNT buckypaper. Moreover, the current practice of quality inspection, mainly based on operators' visual inspection, has three limitations: (1) subjective judgments by operators, (2) requirement of sophisticated training of operators, (3) slow reaction to the alert and lack of capability for real-time quality control. Therefore, a data-driven methodology is needed to perform a real-time quality assessment in a unified manner.

The objective of this paper is to propose a standard real-time quality quantification methodology for the continuous manufacturing of CNT buckypaper, which directly reflects the quality information, such as impurity, alignment, functionalization, thickness, long-term consistency, and uniformity quickly and accurately. This real-time data-driven quality assessment methodology will provide quality control guidance to engineers for the CNT buckypaper industry.

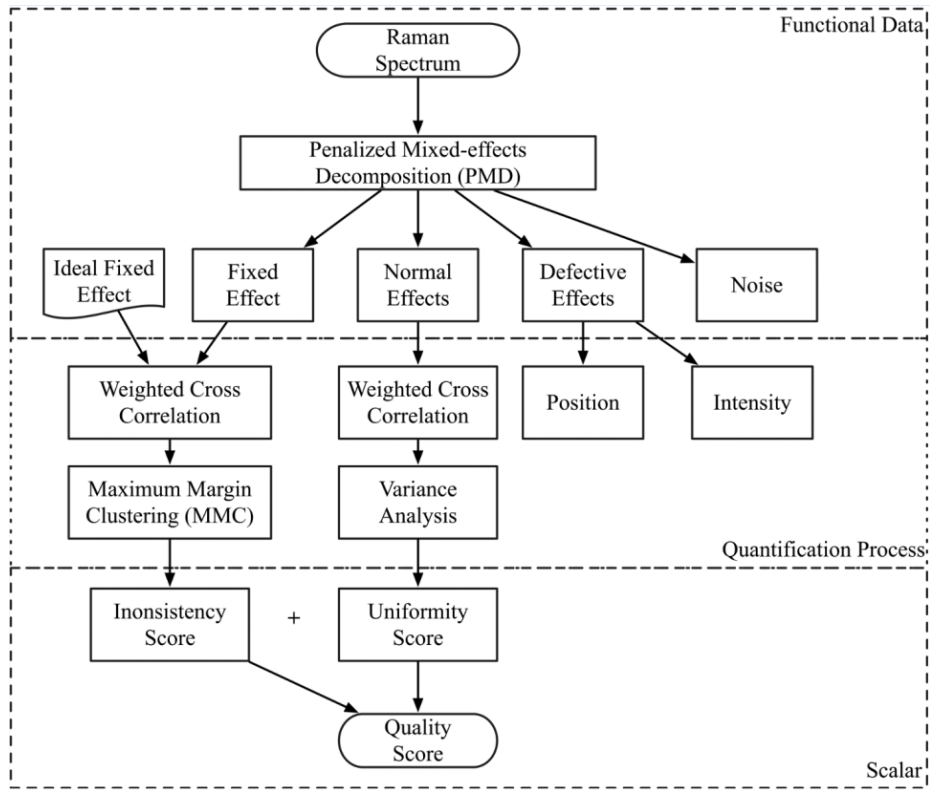

Fig. 2.  In-line Raman Spectroscopy Inspection for Buckypaper Production

As shown in Fig. 2, after collecting in-line Raman spectra, we first apply the PMD algorithm [19] to extract fixed effects, normal effects, and defective effects from Raman spectra. Since one can easily interpret defective effects by the position and the intensity of the corresponding defects, we will focus on the study of the fixed effects and normal effects, which have no standard evaluation criterion. We use the proposed quality quantification approach to measure their between-sample differences (fixed effects) from the ideal sample and the within-sample variation (normal effects). After quantifying these parameters, an overall quality score is proposed to represent the product quality of CNT buckypaper in a unified manner. The obtained quality score has three key characteristics: (1) it should be distinguishable for CNT buckypaper samples with different quality levels, (2) it should be interpretable with corresponding physical features, (3) it should be easily obtainable from Raman spectroscopy inspection. The quality score provides guidance for operators to tackle process issues and improve the quality of CNT buckypaper. Based on the quality score, an operator could quickly determine if there is a process mean shift, local uniformity issue, or some defects. The stakeholders, such as customers, managers can realize intelligent decision making according to real-time data-driven quality assessment.

The remainder of this paper is organized as follows. Section II illustrates a systematic CNT buckypaper quality assessment process. Section III presents a case study to demonstrate the implementation procedures. Finally, a summary is provided in Section IV.

## II.  Real-time Data-driven Quality Assessment

### A.  Penalized Mixed-effects Decomposition (PMD) for In-line Raman Spectroscopy

The in-line Raman spectra are collected and classified into N groups according to a pre-designed maximin Latin Hypercube Design [23]. A wavelet-based penalized mixed-effects decomposition (PMD) [19] is formulated for multichannel profile detection of in-line Raman spectroscopy as

$$\mathbf{y}_{ij} = \boldsymbol{\mu}_i + \mathbf{W}\boldsymbol{\theta}_{ij} + \mathbf{W}_a\boldsymbol{\delta}_{ij} + \mathbf{e}_{ij}, \tag{1}$$

where $\mathbf{y}_{ij}$ is a measurement profile with dimension $n$, corresponding to the $j^{\text{th}}$ profile in the $i^{\text{th}}$ sample; $\boldsymbol{\mu}_i$ denotes fixed effects in the $i^{\text{th}}$ sample; $\mathbf{W}$ and $\mathbf{W}_a$ are wavelet-based design matrices with dimension $n \times p$ and $n \times q$ for normal effects and defective effects respectively; $\boldsymbol{\theta}_{ij}$ and $\boldsymbol{\delta}_{ij}$ are coefficients vectors associated with normal effects and defective effects with regard to the $j^{\text{th}}$ profile in the $i^{\text{th}}$ sample. $\mathbf{e}_{ij}$ represents a signal-dependent noise vector for the $j^{\text{th}}$ profile in the $i^{\text{th}}$ sample. The three decomposed effects are represented as $\boldsymbol{\mu}_i, \mathbf{W}\boldsymbol{\theta}_{ij}$ and $\mathbf{W}_a\boldsymbol{\delta}_{ij}$, which are (i) the *fixed effects*, $\boldsymbol{\mu}_i$, that reveals the fabrication consistency of the sample, i.e., long-term mean shift along the fabrication process, (ii) the *normal effects*, $\mathbf{W}\boldsymbol{\theta}_{ij}$, that quantify the uniformity of quality features in the sample area, and (iii) the *defective effects*, $\mathbf{W}_a\boldsymbol{\delta}_{ij}$, that show the exist of specific sampling points with defective quality features. Therefore, the PMD is one kind of deterministic data decompositions [24] and it could identify the features related to the geometric shape, topological structure, or algebraic characteristics [25].

However, the PMD has the following limitations when it is applied to real-time quality assessment for CNT buckypaper manufacturing process:

(1) It can decompose Raman Spectra into fixed effects, normal effects and defective effects. However, a gap exists between the extracted features and the quality characteristics of continuous fabricated CNT buckypaper (e.g., consistency and uniformity properties).

(2) Although the PMD provides multiple dimensions to reflect CNT buckypaper performance, a single meaningful composite quality index is needed to reflect the overall CNT buckypaper quality.

To overcome those two limitations, we propose to quantify the fixed effects and the normal effects by using weighted cross-correlation to measure the similarity between profiles. The variance analysis is then applied to measure the deviation of the similarity of the normal effects.

### B.  Weighted Cross-correlation for Profiles Similarity Quantification

The fixed effects are driven from the mean vector of multiple profiles in each group of the sample that reflect the long-term mean shift of the fabricating process. This can be measured by the (dis)similarity between the sample's fixed effects and the ideal fixed effects. Similarly, the (dis)similarities among within-sample normal effects represent the degree of within-sample (dis)order.

There are multiple definitions of (dis)similarity measurement, whereas the best selection of a similarity measurement depends on specific domain knowledge. In the research area of the spectral library, an upcoming spectrum is searched among a known spectral library to find the optimal match. This match is captured by the similarity measure. The most common similarity measurements used in spectral library search are Euclidean, Mahalanobis, Pearson correlation coefficient absolute value, citiblock, cosine, and least square. The Euclidean similarity, citiblock, Mahalanobis, and correlation coefficient are classical point-to-point measurements, which are unable to deal with a minor shift and line broadening difference [26]. The dissimilarity will increase significantly due to a small shift in peak positions if point-to-point measurements are used [27]. However, the similarity measures which consider the neighborhoods can quickly capture the minor shifts. One popular similarity measurement which considers the neighborhoods is the weighted cross-correlation based generalized expression of similarity method [27]. Inspired by [27], we propose to apply weighted cross-correlation to quantify the long-term mean shift and the within-sample disorder that may happen in the CNT buckypaper fabrication process.

In a Euclidean vector space $V$, we define the profile as a continuous mapping $\boldsymbol{\beta}_i : [1, n] \rightarrow V$, where n is the dimension of the measurement profile $\mathbf{y}_{ij}, n \in \mathfrak{R}$ and $n \geq 1$. The ideal profile is $\boldsymbol{\beta}_0 : [1, n] \rightarrow V$. Given two profiles $\boldsymbol{\beta}_0 : [1, n] \rightarrow V$ and $\boldsymbol{\beta}_i : [1, n] \rightarrow V$, the cross-correlation function $c_{0i}(r)$ for pattern $\boldsymbol{\beta}_0(x)$ and $\boldsymbol{\beta}_i(x)$ is defined as:

$$c_{0i}(r) = \int \boldsymbol{\beta}_0(x)\boldsymbol{\beta}_i(x + r)dx, \tag{2}$$

where $r$ is the relative shift (lag) between those two functions, $\boldsymbol{\beta}_0(x)$ and $\boldsymbol{\beta}_i(x)$. The similarity between $\boldsymbol{\beta}_0(x)$ and $\boldsymbol{\beta}_i(x)$ is given by

$$S_{0i} = \frac{\int (1 - \frac{|r|}{l}) \mathbb{I}_{\{|r|<l\}} \times c_{0i}(r) dr}{\sqrt{\int (1 - \frac{|r|}{l}) \mathbb{I}_{\{|r|<l\}} c_{00}(r) dr \int (1 - \frac{|r|}{l}) \mathbb{I}_{\{|r|<l\}} c_{ii}(r) dr}}, \qquad (3)$$

where $l$ defines the width of the neighborhoods considered, $c_{00}(r)$ and $c_{ii}(r)$ are the auto-correlation functions that are defined in analogy to Equation (2), $\mathbb{I}_{\{|r|<l\}}$ is the indicator function that gets the value 1 for $|r| < l$, and the value 0 for $|r| \geq l$.

The dissimilarity between the pattern $\boldsymbol{\beta}_0(x)$ and $\boldsymbol{\beta}_i(x)$ is therefore given by:

$$D_i = (S_{00} + S_{ii} - 2S_{0i})/2, \qquad (4)$$

where $S_{00}$ and $S_{ii}$ are the self-similarity of the pattern $\boldsymbol{\beta}_0(x)$ and $\boldsymbol{\beta}_i(x)$ respectively, and $S_{00} = S_{ii} = 1$.

To include the neighborhood into the calculation of the (dis)similarity, one should define the value of r $\neq$ 0. The dissimilarity criterion $D_i$ will yield a value of 1 when the patterns of $\boldsymbol{\beta}_0(x)$ and $\boldsymbol{\beta}_i(x)$ are perfectly dissimilar, a value of 0 when patterns are identical, and a value between 0 and 1 for otherwise.

### C. Formulation for inconsistency index between samples

The inconsistency of a process is the long-term mean shift that happened to the process. Since the long-term mean shift is captured by the fixed effect, the between samples' consistency can be assessed by the dissimilarity changes of the fixed effects. We first adopt weighted cross-correlation to measure the general dissimilarity. The two profiles in this case would be $\boldsymbol{\mu}_i(x)$ and $\boldsymbol{\mu}_0(x)$, where $i = 1, 2, \ldots N$ is the index of the samples, N is the total number of samples, 0 is the index of the ideal profile.

Besides, different types of features should be considered in the index development. For example, Xiang et al. incorporated the jumps and phase variability with the profile change detection [28]. Since the normalization in Equation (3) dilutes the mean shift of peak intensity, we introduce the maximum intensity difference ($d_i = |\max(\boldsymbol{\mu}_0(x)) - \max(\boldsymbol{\mu}_i(x))|$) to the consideration. For a sample $i$, let $\boldsymbol{z}_i^T = (d_i, D_i) \in R^2$ be the row vectors of a collection of data points, arranged as the rows of the matrix $\boldsymbol{Z} \in R^{N \times 2}$. Our main interest is to separate the data into consistent and non-consistent classes in a large margin classifier. Given data $\boldsymbol{z}_1, \boldsymbol{z}_2, \ldots, \boldsymbol{z}_N$, these data points would be assigned into two classes as $\eta_i \in \{-1, +1\}$, arranged as $\boldsymbol{\eta} = (\eta_1, \eta_2, \ldots, \eta_N)^T$, when $\eta_i = -1$, the sample $i$ is consistent with the others, when $\eta_i = 1$, the sample $i$ is inconsistent with the others. In such a way, the separation between two classes is as wide as possible, which is known as unsupervised large margin method.

Unsupervised large margin methods, notably the maximum margin clustering (MMC) [29] is a popular clustering method that is motivated by the support vector machines (SVM). Without loss of generality, we assume the data set has been standardized as in the general procedure of MMC. Mathematically, the MMC approach aims at solving the following optimization problem [29]:

$$\min_{\boldsymbol{\eta}} \min_{\boldsymbol{w},b} \left\| \boldsymbol{w} \right\|^2 + 2C \boldsymbol{\xi}^T \boldsymbol{e} \qquad (5)$$
$$\text{s.t. } \eta_i(\boldsymbol{w}^T \boldsymbol{z}_i + b) \geq 1 - \xi_i, \ \xi_i \geq 0,$$
$$\eta_i = \{\pm 1\}, -\ell \leq \boldsymbol{e}^T \boldsymbol{\eta} \leq \ell,$$

where $\boldsymbol{\xi} = [\xi_1, \ldots, \xi_N]^T$ is the vector of slack variables ($\xi_i, i = 1, \ldots, N$) for the errors, C > 0 is a regularization parameter and $\boldsymbol{e}$ is the vector of ones, and $\ell \geq 0$ is a constant controlling the class imbalance. This optimization problem can be solved by using the iterative approach [30].

The distance from the data points to the optimal decision surface is used as the decision value in MMC. The confidence level of the probability for predicting true class increases when the decision value is large. The decision value, therefore, is an indicator for labeling consistent and inconsistent samples. To develop a single composite index for buckypaper consistency assessment, we adopt the decision value in MMC. For the purpose of buckypaper consistency assessment, the desired direction (e.g., small dissimilarity and maximum intensity difference) is already known. However, the signs of the decision value can be misleading to the engineers' intuition. To address this problem, we transfer the decision value by an arbitrary function as described in Equation (6). The advantages of this transformation are: (1) when the shape parameter $\rho > 1$, the function is monotonic; (2) the interpretability of the index will be improved as the index would become non-negative in this study; (3) the sensitivity of the index will be improved as the Weibull cumulative distribution function will have a sharp increase at the boundary of shifting, and (4) the transform is invertible.

In our case, since $\boldsymbol{z}_i^T \in R^2$ and $\boldsymbol{w}^T = (w_1, w_2) \in R^2$, the decision surface is $\boldsymbol{w}^T \boldsymbol{z}_i + b = 0$. The decision value is $\boldsymbol{\tau} = (\tau_1, \tau_2 \ldots, \tau_N)^T$, where $\tau_i = \frac{|\boldsymbol{w}^T \boldsymbol{z}_i + b|}{\|\boldsymbol{w}\|_2}$.

The inconsistency index is therefore defined as:

$$C_i = 1 - e^{-\left(\frac{-\eta_i \tau_i - \min(-\boldsymbol{\eta} \circ \boldsymbol{\tau})}{\lambda}\right)^\rho}, \qquad (6)$$

where $\rho > 1$ is the *shape parameter* and $\lambda > 0$ is the *scale parameter* that needs further calibration according to the domain knowledge and the in-control data of a specific CNT buckypaper product. $\tau_i$ is the distance from a sample $i$ to the decision surface, while $\eta_i$ is the clusters that the sample belongs to (when $\eta_i = -1$, the sample $i$ is consistent with the others, when $\eta_i = 1$, the sample $i$ is inconsistent with the others). The elementwise matrix product of $\boldsymbol{\eta} \in R^N$ and $\boldsymbol{\tau} \in R^N$ is denoted by $\boldsymbol{\eta} \circ \boldsymbol{\tau}$, i.e., the Hadamard product.

The threshold of the inconsistency score is the value at $\tau_i = 0$, which means $\boldsymbol{z_i}$ is on the decision surface, and we further transform it to the inconsistency index space as

$$\Delta = 1 - e^{-\left(\frac{-\min(-\boldsymbol{\eta} \circ \boldsymbol{\tau})}{\lambda}\right)^\rho},$$

$C_i$ reflects the changes of fabrication consistency due to the long-term process mean shift. The inconsistency index will be a value equal to 0 when the ideal fixed effects and the sample fixed effects are identical; otherwise, it would be a scaled number between zero and one. When $C_i \geq \Delta$, the sample $i$ is inconsistent with other samples; otherwise, the sample is consistent with the others. The calculation steps are illustrated in the Algorithm 1.

| **Algorithm 1** Inconsistency index procedure |
|---|
| 1: Input the dissimilarity $D_i$ as described in section II. B. and the maximum intensity difference $d_i$. |
| 2: Initialize the labels $\boldsymbol{\eta}$ by simple clustering method. |
| 3: Fix $\boldsymbol{\eta}$ and train standard SVM model. |
| 4: Compute the $\boldsymbol{w}$ and b from the KKT conditions. |
| 5: Assign the labels as $\eta_i = sign(\boldsymbol{w^T z_i} + b)$. |
| 6: Repeat 3-5 until convergence. |
| 7: Return the labels $\boldsymbol{\eta}, \boldsymbol{w}$ and b. |
| 8: Compute the decision values $\tau_i = \frac{|\boldsymbol{w^T z_i} + b|}{\|\boldsymbol{w}\|_2}$. |
| 9: Return the inconsistency score $C_i$ from Equation (6). |

### D. Formulation for uniformity index within samples

The normal effects provide us the information that relevant to the degree of alignment, the degree of functionalization, nanotube distribution, and dispersion of the CNT buckypaper sample. The uniformity between the normal effects in each sample reveals relatively robust performance on alignment, functionalization, distribution, and dispersion.

The normal effects of sample $i$ is denoted as $\boldsymbol{\theta}_{ij}(x)$, where $j = 1, \dots, n$ is the index of normal effects within the sample. Two functions $\boldsymbol{\theta}_{ij}(x)$ and $\boldsymbol{\theta}_{ik}(x)$ are the normal effect functions in sample $i$, where $x$ is the wavelength index. The similarity between $\boldsymbol{\theta}_{ij}(x)$ and $\boldsymbol{\theta}_{ik}(x)$ in sample $i$ is defined as $S_{jk}^i$, while the cross-correlation, auto-correlation are defined as $c_{jk}^i(r)$, $c_{jj}^i(r)$ and $c_{kk}^i(r)$.

Within sample $i$, the similarity between observations $j$ and $k$ is known as $S_{jk}^i$. This similarity criterion will yield a value of 1 when the patterns of $\boldsymbol{\theta}_{ij}(x)$ and $\boldsymbol{\theta}_{ik}(x)$ are identical, and a value between 0 and 1 for other cases. However, this criterion cannot directly reflect the within-sample uniformity as there will have $n \times n$ similarity matrix for a sample.

From the statistical perspective, the uniformity implies variability among mutual similarities of the observations within one sample. The uniformity within one sample $i$ is then defined as:

$$U_i = \frac{\sum_{j=1}^n \sqrt{\frac{1}{n-1} \sum_{k=1}^n \left(S_{jk}^i - \bar{S}_{j.}^i\right)^2}}{n}, \quad (7)$$

where $S_{jk}^i$ is the similarity between observations $j$ and $k$ in sample $i$, and $\bar{S}_{j.}^i = \sum_{k=1}^n S_{jk}^i / n$.

This index indicates the uniformity disorder of sample $i$ due to within-sample random variations. The uniformity quantification criterion yields a scaled value from zero to one. A lower value of this index shows that the normal effects within the sample $i$ tend to have better uniformity.

### E. Overall Quality Quantification and Interpretation

To quickly check and rank the CNT buckypaper quality, a composed index is needed. Since the defective effects directly reflect the quality issue, we further use the consistency index ($C_i$), and the uniformity index ($U_i$) to compose the total quality index of the CNT buckypaper sample $i$:

$$Q_i = W_1 C_i + (1 - W_1) U_i, \quad (8)$$

where $W_1$ and $1 - W_1$ are the weights of the consistency and uniformity indices. These weights are chosen based on the significance level from engineering domain knowledge.

The total quality index yields zero when the process is consistently, uniformly producing CNT buckypaper that are identical to the designed product. Otherwise, it would be a value between zero and one to indicate the quality status of the samples. This single meaningful composite quality index is desirable to provide a quick and effective overall quality performance assessment of a CNT buckypaper manufacturing process.

## III. Case Study

### A. Experiment Preparation and Raman Spectra Interpretation

The fabrication detail of the CNT buckypaper with random alignment can be found in [11], [31]. The CNT buckypapers' typical thickness in this experiment was 10 μm ~ 20 μm. This thickness is measured by using Heidenhain-Metro incremental length gauge, and further confirmed by a SEM measurement. Fig. 3 shows the in-line Raman spectroscopy inspection system. Fig. 3 (a) is the overlook of the Renishaw™ Invia Micro-Raman System. In the experimental set-up of this study, as shown in Fig. 3 (b), we have a custom-designed remote optical probe and roller sample stage. For the remote probe, near-infrared (NIR) laser with a wavelength of 785 nm and a power of 150 mW were used to eliminate the effect of ambient lights. Low magnification lens was used to achieve a more considerable focus tolerance. Fig. 3 (3) shows that for each sample on the roller, the inspection system will measure the Raman spectra according to the per-determined design of experiment. The details about data collection will be discussed in the next subsection.

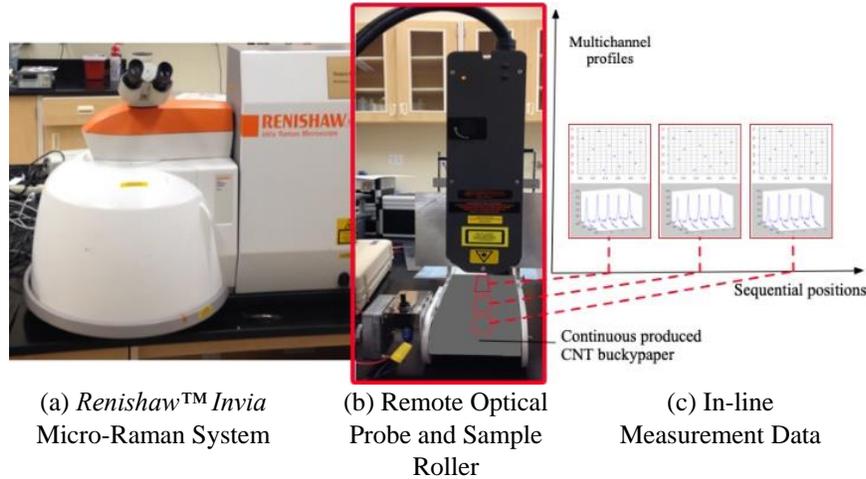

(a) *Renishaw™ Invia* Micro-Raman System    (b) Remote Optical Probe and Sample Roller    (c) In-line Measurement Data

Fig. 3. *Renishaw™ Invia* Micro-raman System with Custom-Designed Remote Optical Probe and Sample Roller for In-line Measurement.

Raman spectra are used for detecting the quality information of the CNT buckypaper. It provides us with information relevant to potential quality issues such as impurity, the degree of chemical functionalization, and alignment of CNTs in buckypaper. It is known that Radial Breathing Mode ($< 300 \ cm^{-1}$) is used to determine the diameter of single-walled carbon nanotube (SWCNT); D-band ($1250 \sim 1400 \ cm^{-1}$) is related to the disorder or molecular defects in the CNT structure; and D-band to G-band intensity ratio is an effective indicator of CNT quality of functionalization. In addition, polarized Raman provides angular dependence of the Raman intensity, and then the degree of CNT alignment can be estimated [1].

### B. Data Collection

Fig. 4 provides a further illustration of the design of experiment and the collected Raman Spectra for the CNT buckypaper samples on the roller. Within each sample, we use a Maximum Latin Hypercube Design, as shown in Fig. 4 (a), to pre-determine the positions of a certain number of Raman spectra that needs to be collected in a unit square. The design of experiment of this kind has a good space-filling property and the first-dimension projection property. The Raman spectra, as shown in Fig. 4 (b), located at the corresponding sample points, are collected for inspecting the quality of the CNT buckypaper. These Raman spectra are used to extract a quantitative index to represent the quality in the sample area.

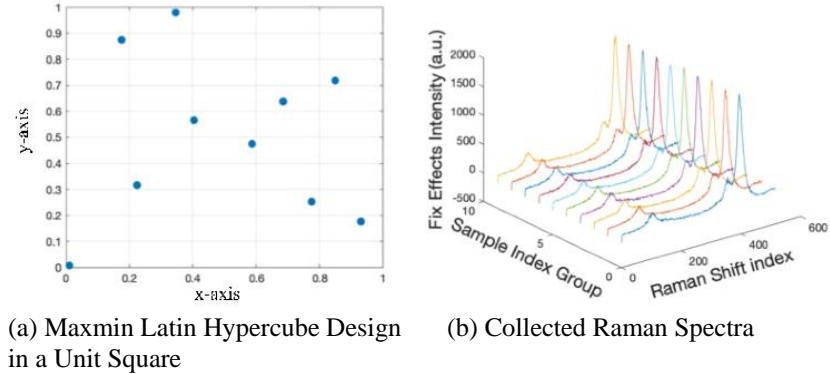

(a) Maxmin Latin Hypercube Design in a Unit Square     (b) Collected Raman Spectra

Fig. 4. The Design of Experiment and the Corresponding Raman Spectra

In our experiment, the Raman spectra are collected in the measurement zone with a rectangular shape with $120 \times 120$ micrometers. We collect ten samples, and within each of them, ten observation points are collected. These observations are tested sequentially and a Raman spectrum with 512 Raman shifts and intensities is collected for each measurement point. All the Raman spectra are collected based on a piece of single-wall CNT buckypaper. In the *Renishaw™ Invia* Micro-Raman System, Raman Microscopy with 785 nm laser source and 0.5-second exposure time for each measurement point is conducted.

If each representative sample can be regarded as a sensor channel to collect Raman spectra, the process modeling and detection for the CNT buckypaper fabrication process can be formulated as a multichannel profile modeling problem along with the sequential position of the CNT buckypaper. Thus, we use the PMD to process the collected Raman spectra.

TABLE I.
LONG-TERM MEAN SHIFT DETECTION COMPARISON AMONG DISSIMILARITY, MAXIMUM INTENSITY DIFFERENCE AND INCONSISTENCY (BOLD AND UNDERLINED NUMBER ARE BEYOND LIMITS)

| | | | *Dissimilarity* | | | | | | *Maximum Intensity Difference* | | | | *Inconsistency\** | |
| | | CUSUM ($h = 5$) | | EWMA ($\lambda = 0.2$) | | | | CUSUM ($h = 5$) | | EWMA ($\lambda = 0.2$) | | | Proposed Method | |
| Sample Index | $D$ | $C_i^+$ | $C_i^-$ | $z_i$ | LCL | UCL | $d$ | $C_i^+$ | $C_i^-$ | $z_i$ | LCL | UCL | $c_i$ | UCL |
|---|---|---|---|---|---|---|---|---|---|---|---|---|---|---|
| 1 | 0 | 0 | -1.85671 | 0.000104 | 9.69E-05 | 0.000163 | 0 | 0 | -1.8325 | 89.0412 | 82.67089 | 139.9321 | 0 | 0.2899 |
| 2 | 0.00002 | 0 | -3.35086 | **8.72E-05** | 8.76E-05 | 0.000172 | 65.55 | 0 | -2.2913 | 84.34296 | 74.63642 | 147.9666 | 0.00197 | 0.2899 |
| 3 | 0.00008 | 0 | -3.75729 | 8.58E-05 | 8.26E-05 | 0.000177 | 89.96 | 0 | -2.23854 | 85.46637 | 70.3127 | 152.2903 | 0.00691 | 0.2899 |
| 4 | 0.00016 | 0.043857 | -2.71343 | 0.000101 | 7.97E-05 | 0.00018 | 98.285 | 0 | -2.01132 | 88.03009 | 67.77031 | 154.8327 | 0.00691 | 0.2899 |
| 5 | 0.00016 | 0.087714 | -1.66957 | 0.000112 | 7.79E-05 | 0.000182 | 84.605 | 0 | -2.07079 | 87.34508 | 66.21836 | 156.3846 | 0.00277 | 0.2899 |
| 6 | 0.00008 | 0 | -2.076 | 0.000106 | 7.68E-05 | 0.000183 | 69.675 | 0 | -2.44314 | 83.81106 | 65.25256 | 157.3504 | 0.00166 | 0.2899 |
| 7 | 0.00022 | 1.131571 | 0 | 0.000129 | 7.61E-05 | 0.000184 | 54.745 | 0 | -3.12837 | 77.99785 | 64.64493 | 157.9581 | 0.00006 | 0.2899 |
| 8 | 0.0002 | 1.900571 | 0 | 0.000143 | 7.56E-05 | 0.000184 | 40.205 | 0 | -4.11831 | 70.43928 | 64.26017 | 158.3428 | 0.00001 | 0.2899 |
| 9 | 0.00014 | 1.581857 | 0 | 0.000142 | 7.53E-05 | 0.000185 | 241.72 | 2.233127 | -0.88518 | 104.6954 | 64.01556 | 158.5874 | **0.68565** | 0.2899 |
| 10 | 0.00024 | 3.076 | 0 | 0.000162 | 7.52E-05 | 0.000185 | 368.27 | **7.11831** | 0 | 157.4103 | 63.85968 | 158.7433 | **0.99989** | 0.2899 |

*C. Results and Discussions*

After processing by the PMD, the Raman spectra are decomposed into fixed effects, normal effects, defective effects, and signal dependence noise. Fig. 5 compares the fixed effect with the corresponding inconsistency index.

Fig. 5 (a) shows the comparison between ideal Raman spectra's fixed effect and the samples' fixed effects. The $0^{th}$ sample is the ideal fixed effect, while the $1^{st}$ to $10^{th}$ samples are the fixed effects decomposed from the real data, as shown in Fig. 5 (b). The shape parameter and the scale parameter are selected to be 5 and 2, respectively, in Equation (6). Via empirical study based on in-control testing datasets, we find the inconsistency function has a sharp slope when the shape parameter is equal to 5. When the scale parameter is equal to 2, the minimum value and the maximum value of the slope are at the margin, which separates the samples into the consistent and inconsistent groups.

Table I compares the long-term mean shift detection of the dissimilarity, maximum intensity difference and the proposed inconsistency. By using dissimilarity alone, Exponentially Weighted Moving Average (EWMA) detects the $2^{nd}$ sample beyond the limits. The Cumulative Sum (CUSUM) control chart for the maximum intensity difference detects the $10^{th}$ sample beyond the limit. The proposed method detects that the $9^{th}$ and $10^{th}$ samples have mean shifts, which is identical to the underlying true observation shown in Fig. 5 (a). Overall, the samples have a long-term mean shift comparing with the ideal fixed effects, especially from the $9^{th}$ and $10^{th}$ samples.

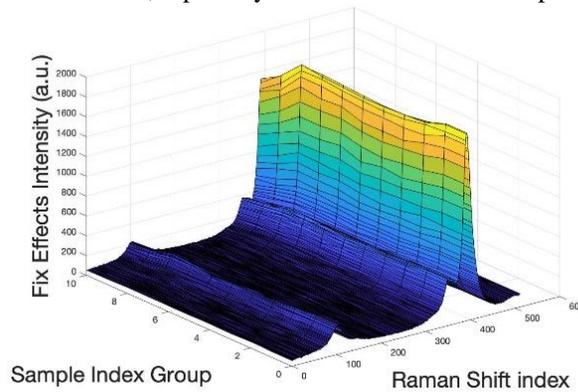

(a) Fixed Effects Separated by the PMD in Reference Ideal Raman Spectrum (0) and Real-Data Raman Sample (1-10)

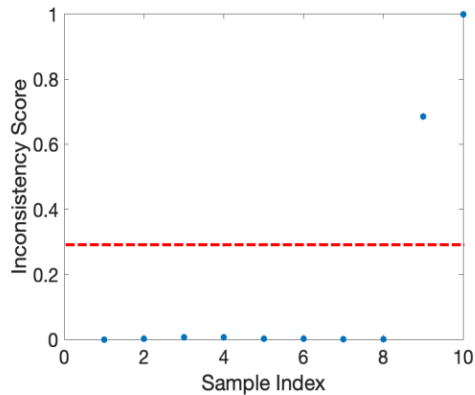

(b) the Relative inconsistency Score

Fig. 5.  Comparison Between Fixed Effects and the Corresponding Inconsistency Score

The inconsistency index is implemented to capture the long-term mean shift of the real data sample, as shown in Fig. 5 (b). The consistency rank of the samples from the good to the bad is the order of the $1^{st}$, $8^{th}$, $7^{th}$, $6^{th}$, $2^{nd}$, $5^{th}$, $4^{th}$, $3^{rd}$, $9^{th}$, and $10^{th}$ samples. The Raman intensity after the $9^{th}$ sample tends to be larger than the previous samples and the ideal sample. This change may result from the measurement equipment as the focus depth changes due to the sample local deformation. Another possible reason could be process changes. Some unknown process changes shift the process mean accidentally so that the $9^{th}$ sample starts to be inconsistent with the other samples.

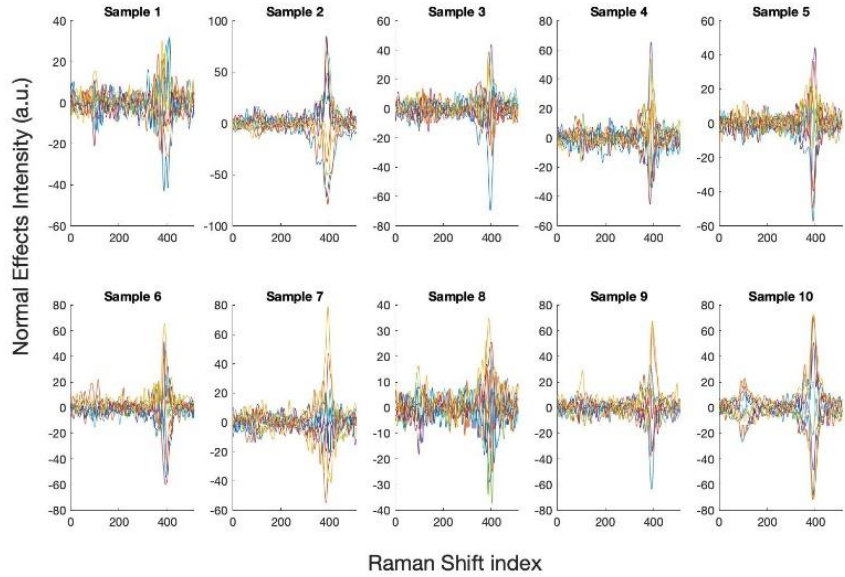

(a) the Normal Effects of the Real Data

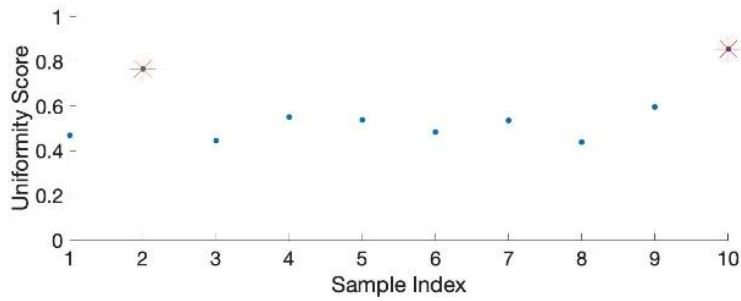

(b) the Corresponding Uniformity Score

Fig. 6.  Comparison between Normal Effects and the Corresponding Uniformity Score

The normal effects of the real data reflect the degree of alignment, the degree of functionalization, nanotube distribution, and dispersion of the sample at the profile level. Fig. 6 compares the normal effects and the corresponding uniformity index proposed in this paper. As shown in Fig. 6 (a), it is hard to observe the entropy of the normal effects within the sample. Therefore, the uniformity index is calculated, and the corresponding uniformity index for the samples are shown in Fig. 6 (b). The width of the neighborhood considered $l$ is 2. The rank of the samples' uniformity from the good to bad is the 8th, 3rd, 1st, 6th, 7th, 5th, 4th, 9th, 2nd, 10th sample. The uniformity of the 2nd and 10th samples reflect the product quality issues, such as the degree of functionalization, the degree of alignment changes, nanotube distribution and dispersion.

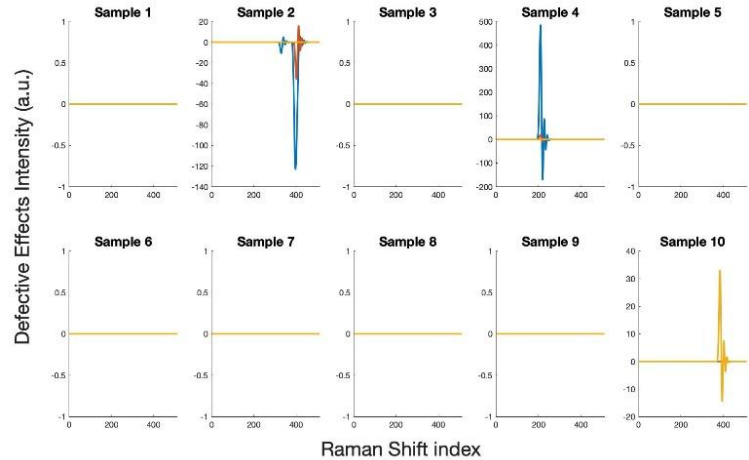

Fig. 7. The Defective Effects of the Real Data

Fig. 7 shows the decomposed defective effects of the read data. After the PMD extracts defective effects from the Raman spectra, one can observe that the observations #1 and #2 in the $2^{nd}$ sample, observations #8 and #9 in the $4^{th}$ sample, and observations #9 in the $10^{th}$ sample are defective. Based on the Raman shift, one can find that the defective effects of the 2nd and 10th samples occur at the G-band. The defects are due to the impurity of raw material, different degrees of functionalization, different alignments of carbon nanotubes, or bad nanotube dispersion. These quality issues can also be reflected in the uniformity quantification. The defective effect of the 4th sample is located between the D-band and the G-band, and it might result from some measurement errors, such as external light disturbance.

The defective effects of the $2^{nd}$ and $10^{th}$ samples occur in the G-band. The defects are due to the impurity of raw material, different degrees of functionalization, different alignments of carbon nanotubes, or bad nanotube dispersion. These quality issues can also be reflected in the uniformity quantification. The defective effect of the $4^{th}$ sample is located between the D-band and the G-band, and it might result from some measurement errors, such as external light disturbance. One can keep the $4^{th}$ sample in mind, and we will discuss this measurement error later. Fig. 8 shows the overall quality performance of those ten samples ($w_1 = 0.3$). The overall quality rank of the samples from the good to bad is the $8^{th}$, $3^{rd}$, $1^{st}$, $6^{th}$, $7^{th}$, $5^{th}$, $4^{th}$, $2^{nd}$, $9^{th}$, $10^{th}$ sample.

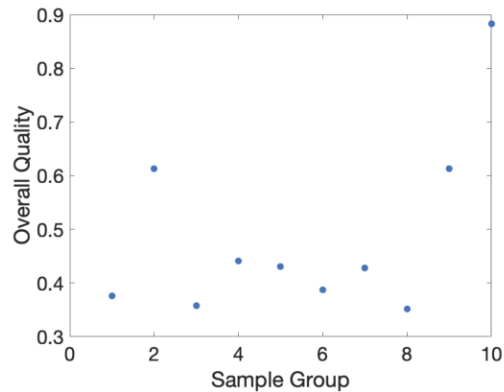

Fig. 8. The Overall Quality of the Real Data

Although the overall quality index does not consider the defective effect, one can find that the samples with defects will be identified as not qualified samples by the proposed index. From the overall quality values, we could see that the $4^{th}$ sample has a relatively smaller value than other defective samples. The quality of the $4^{th}$ sample is comparable with samples without defective effects. In [19], the authors suspect the observed defective phenomenon on the $4^{th}$ sample results from some measurement errors, such as external light. Our overall quality index helps to verify that the defective effect of the $4^{th}$ sample is caused by measurement errors.

TABLE II.
THE OVERALL QUALITY OF THREE BUCKYPAPER MATERIALS

| Sample ID<br>Material | 1 | 2 | 3 | 4 | 5 |
|---|---|---|---|---|---|
| Raw SWCNT | 0.40781 | **0.60958** | 0.40624 | 0.38501 | **0.59167** |
| Acid SWCNT | 0.3912 | **0.63058** | 0.42267 | 0.48447 | **0.59869** |
| Functionalized SWCNT | 0.40339 | **0.57911** | 0.40266 | **0.57829** | 0.43248 |

We further apply the proposed quality assessment method to three buckypaper materials. They are raw SWCNT, SWCNT after acid, and SWCNT after functionalization. Table II summarizes the overall quality score for all the samples of the materials. The threshold under the case study setting is 0.5. Therefore, the samples 2 and 5 of raw SWCNT and acid SWCNT have relatively low quality, while the samples 2 and 4 of functionalized SWCNT have relatively poor quality as their overall quality scores are beyond 0.5. This result is matching up with the conclusion by the experienced engineer.

## IV. Conclusions And Future Work

In the continuous CNT buckypaper manufacturing process, a complicated profile data, called in-line Raman spectroscopy, is used to collect Raman spectra for the CNT buckypaper quality monitoring. The Raman spectra fuse affluent information that includes quality consistency, local uniformity, and within-sample defects. The PMD method enables us to extract fixed effects, normal effects, and defective effects from the Raman spectra. Although multiple quality features are decomposed from the Raman spectra, these features cannot be used to evaluate the real-time fabrication quality of CNT buckypaper directly. Current practice relies on heuristic methods based on these quality features, which has specific limitations, including (i) subjective judgments by operators, (ii) requirement of sophisticated training of operators, and (iii) slow reaction to the process changes. It is important to develop novel quality assessment indices for the system to automatically evaluate the product quality in a real-time manner.

The main contribution of this paper is to propose a new real-time quality assessment index to access the quality characteristics of samples based on in-line Raman spectra in a continuous CNT buckypaper manufacturing process. The proposed quality assessment indices quantify the CNT buckypaper quality from fixed effects and normal effects. The inconsistency index derived from fixed effects reveals the long-term mean shift of the process, while the uniformity index originated in normal effects reflects the within-sample uniformity. The overall quality index considers both uniformity and consistency to evaluate the quality of the CNT buckypaper. All these three indices yield from zero to one to show the corresponding quality characteristics from good condition to bad condition. In the case study, the proposed assessment approach is applied to distinguish the quality performance of the different CNT buckypaper samples. The proposed indices successfully identify the long-term mean shift that occurs in the process as well as the samples with the large within-sample disorder. Also, our proposed approach can provide quantitative quality indices for single-walled carbon nanotube buckypaper after acid processing or functionalization. The quality assessment results are consistent with evaluations from the experienced engineers. The interpretation based on the proposed quantification indices is clear to the corresponding physical features. It is not only obtainable from Raman spectroscopy inspection but also controllable for operators to tackle process issues. By applying the proposed assessment method, the quality of CNT buckypaper could be quantified from the consistency, uniformity, and defective perspectives.

In future work, the correlation between the proposed indices and the control variables of the CNT buckypaper fabricating process will be studied. Systematic in-line quality assessment and process improvement methodologies will be explored.

## SUPPLEMENTARY MATERIALS

The Raman spectra based on raw SWCNT buckypaper, acid- treated SWCNT buckypaper, and functionalized SWCNT buckypaper are provided in the dataset.